# State Based Service Description

*Barbara Paech, Bernhard Rumpe*[*]
*Fakultät für Informatik,*
*Technische Universität München*
*80995 Munich, Germany,*
*http://www4.informatik.tu-muenchen.de/*

**Abstract**

In this paper we propose $I/O^*$-state transition diagrams for service description. In contrast to other techniques like for example Statecharts we allow to model non-atomic services by sequences of transitions. This is especially important in a distributed system where concurrent service invocation cannot be prohibited. We give a mathematical model of object behaviour based on concurrent and sequential messages. Then we give a precise semantics of the service descriptions in terms of the mathematical model.

**Keywords**

Semantics, Formal Specification, Service Description, State Transition Diagrams, Mathematical System Model

## 1 INTRODUCTION

The object-oriented paradigm is based on the encapsulation of data within objects. This data can only be accessed by other objects through **service calls**. We use the term **service** as a synonym for **method**. Thus, services are the major constituent for object behavior. However, looking at the different object-oriented analysis and design methods, the abstract specification techniques of services and the interplay between different services within one object still lack a precise semantics. In most cases (e.g. OMT (Rumbaugh, Blaha, Premerlani, Eddy & Lorensen 1991), UML (Booch, Rumbaugh & Jacobson 1997), Syntropy (Cook & Daniels 1994)) state transition diagrams (STD) - inspired by

[*]This paper originated in the SysLab project, which is supported by the DFG under the Leibnizpreis and by Siemens-Nixdorf.

Harels' Statecharts (Harel 1987, Harel & Gery 1996) - are used to specify the object behavior. The STD determines the sequences of object states resulting from service executions. However, services are often not atomic, since even in sequential systems service execution may involve another service execution on the same object. In distributed systems, regarding complex services which involve calls to other objects as atomic, is, in general, a too strong restriction. Objects should react concurrently to as many service calls as possible, while preserving data consistency.

Therefore, we propose to use a whole state transition diagram for the description of one service. Transitions correspond to service steps between an input and an output. Object behavior is derived from the service description by interleaving of the service steps. The service description can also be marked to indicate at which execution states interleaving of other services is allowed.

Because the details of the object behavior are quite intricate, we give a mathematical semantics to object behavior based on the framework of stream processing functions (Broy, Dederichs, Dendorfer, Fuchs, Gritzner & Weber 1993, Klein, Rumpe & Broy 1996) and $I/O^*$-state machines (Rumpe & Klein 1996). In particular, we distinguish sequential and concurrent services calls. This allows to define multiple threads as in Java. As we will show, sequential and purely asynchronous systems are special cases of this model.

Altogether, the paper is structured as follows: First, we introduce the used formal foundation, in particular state machines for the modeling of object behavior. In the following section, we show how to adapt this model to the above sketched communication paradigm. Then we introduce $I/O^*$-state transition diagrams as the abstract description technique for services. We show how to give semantics to object behavior based on the service descriptions.

## 2 MATHEMATICAL SYSTEM MODEL

In (Klein et al. 1996) we developed a formal model of distributed systems, based on the theory of streams (Broy et al. 1993). This mathematical system model serves as a semantical basis for several description techniques, like object models, state transition diagrams, or process diagrams, as for example given in UML (Booch et al. 1997, Breu, Hinkel, Hofmann, Klein, Paech, Rumpe & Thurner 1997).

In this section, we extend the mathematical system model to service descriptions. The model emerged from (Grosu & Rumpe 1995, Rumpe & Klein 1996, Rumpe 1996) where the underlying theory of state machines is developed. In (Grosu & Rumpe 1995) a composition of object behavior is defined.

*Basic assumptions*

We make three basic assumptions about the kind of systems we take into account: First, objects can only read or modify parts of the state of another object through services, even those from the same class. Second, we do not allow more than one service to be active at the same time (however, they may be interleaved). And third, communication between objects is asynchronous such that messages must be accepted, but may be delayed (sequential programming languages correspond to the special case where only one object is active at a time and activity is transferred with service calls).

*$I/O^*$-State machines*

In the following, we introduce the mathematical basis for state based object behavior description. An $I/O^*$**-state machine*** $(S, I, O, \delta, S^0)$ consists of a nonempty set of object states $S$, a nonempty set of input messages $I$, a nonempty set of output messages $O$, a transition relation $\delta \subseteq S \times I \times S \times O^*$, and a nonempty set of initial states $S^0 \subseteq S$.

None of the above given sets need to be finite. The sets of input and output messages $I$ contain service calls and return messages, possibly with arguments. The reaction to any input is attached to the same transition. This leads to a more compact notation compared to the well-known $I/O$-automata* (Lynch & Stark 1989). The transition relation $\delta$ is allowed to be **nondeterministic**. On one hand, this is adequate for the nondeterminism inherent in distributed systems. On the other hand, nondeterminism is important to cope with **underspecification** allowing refinement of such specifications. In (Rumpe 1996, Rumpe & Klein 1996), a refinement calculus for state machines is given which defines a set of development steps to be used for specialization of object behavior during development as well as for inheritance from superclass to subclass. Because of the basic assumptions about systems, an object cannot reject a message. This corresponds to **input enabledness** of the state machine: For each source state $s$ and input message $i \in I$, there exists at least one destination state $t$ and reaction $o \in O^*$ with $\delta(s, i, t, o)$.

*Messages and States*

Object states are composed of several parts that deal with the attribute state and active or suspended service states. We assume that local variables as well as arguments are private to the service invocation they belong to.

Let the set of variables $VAR$ and the set of corresponding values $VAL$ be given. We abstract from the fact that variables are typed, and regard each partial mapping $VAR \rightharpoonup VAL$ as **variable assignment**. We assume that each object has a fixed set of attributes and each service a fixed set of local variables, but do not formalize these constraints here. Given an abstract set $PC$ of **program counters**, suspended service invocations are formalized as $SI = (VAR \rightharpoonup VAL) \times PC \times ID$, where the first component contains arguments and local variables. $PC$ is used to denote special locations in the service code, where a message is awaited and therefore computation is suspended. The third component $ID$ denotes the caller of the service. This is the object, where a

*We call them $I/O^*$**-state machines**, because each transition is labeled accordingly.
*In our classification $I/O$ automata would be called $I \cup O$ automata.

(possible) response is to be delivered. To handle recursion of service calls, as usual, a **stack** of service invocations is used. We assume the mathematical datatype $stack(M)$ over set $M$ with the services $push$, $pop$, $top$ and $\emptyset$ for the empty stack to be given.

If considering multiple threads, one stack is not enough. Indeed, we need a separate stack for each thread. We abstract from actual threads by the set $TAG$, each tag denoting a thread identifier. We incorporate a mapping $TAG \to stack(SI)$ into each object state. Messages are tagged also with elements of $TAG$ to indicate the thread they belong to. Thus, a message is a tuple

$$(sen, rec, tt, mn, ar) \in ID \times ID \times TAG \times MSG \times (VAR \rightharpoondown VAL),$$

where $sen$ is the sender identifier, $rec$ is the receiver identifier, $tt$ is the (thread) tag, $mn$ is the message name, and $ar$ is the argument assignment.
The set $MSG$ contains the service names, but also a special message $ret$ that indicates return messages. The return value (if one exists) is encoded in the arguments of the return message. We use a pool for thread tags for each object, which is used whenever a new thread is started. Each two pools of different objects are disjoint. The states of objects are

$$(at, st, po, pt) \in (VAR \rightharpoondown VAL) \times (TAG \to stack(SI)) \times \mathbb{P}(TAG),$$

where $at$ is the attribute assignment, $st$ is a mapping, which assigns a stack to each thread, and $pt$ is the pool for tags. This set of states is usually infinite. Note that one can easily extend this model to object creation with an additional pool for object identifiers such that object creation is just treated as a special message.

*Transitions*
To model data encapsulation, there are a number of restrictions on the state changes. We shortly repeat the most important restrictions here, without giving a formal definition. The set of attributes of an object and the value of attribute $self$ are immutable. The tag pool may only be diminished. No tag may be used unless removed from the pool. Only one stack is changed in a transition. Either a service invocation is added, removed or the top invocation changed. If the top one is changed, the set of arguments and their values are immutable. Only call messages can add stack elements.

So each transition of the state machine resembles a part of a service execution. If a service calls other services, awaiting their answers, it is partitioned into several transitions.

## 3 MULTI-THREAD COMMUNICATION

In this section, we specialize the behavior model given above to a particular model of communication allowing for service calls where activity is transferred

| $mn_i$ | $mn_o$ | $st'(tt_i)$ |
|---|---|---|
| $sequ$ | $= ret$ | $= st(tt_i)$ |
| $sequ, conc$ | $sequ$ | $= push(st(tt_i), (ar_i + loc, pc, snd_i))$ |
| $conc$ | $conc$ | $= st(tt_i)$ |
| $= ret$ | $= ret$ | $= pop(st)$ |
| $= ret$ | $sequ$ | $= push(pop(st(tt_i)), (ar_i + loc, pc, snd_i))$ |
| $= ret$ | $conc$ | $= st(tt_i)$ |

**Figure 1** Restrictions on $I/O^*$-state machines

(**sequential**) as well as for service calls starting a new thread (**concurrent**). This model could be specialized to purely sequential calls, as in pure C++, or purely concurrent calls. The mixed style presented here is supported in Java, and also is the most flexible for modelling purposes.

Java allows different threads to simultaneously work on the same object and therefore allows to share data. It supports synchronization concepts, but the programmer is responsible to use them correctly. We prevent shared data access by interleaving the service executions. We therefore restrict the Java programming model at this point. However, this can easily be implemented in Java using semaphores. Altogether, we distinguish between **sequential call messages** where the caller awaits the return message, **return messages** that are answers to sequential calls, and **concurrent call messages** that invoke a new thread of computation.

We assume, that no service can compute internally for ever, such that each message is processed. As discussed in (Klein et al. 1996), the communication medium of the general system model ensures that the order of messages is preserved and that message contents are not changed.

Assume a transition $\delta(s, i, t, o)$. Let $s = (at, st, pt)$ be the source state, $t = (at', st', pt')$ the destination state, $i = (snd_i, rec_i, tt_i, mn_i, ar_i)$ the input message and $o = o_1 {+\!+} \langle (snd_o, rec_o, tt_o, mn_o, ar_o) \rangle$ the sequence of output messages, where the last message plays a special role. Only the stack of the input tag $tt_i$ may be changed. Attribute assignments may change arbitrarily. For each concurrent output message in $o_1$ a new tag identifier is removed from $pt$. Sending a concurrent message does not interrupt the active service, but sending of a sequential one does. So only the last message emitted during a transition can be sequential. The tag of a possibly emitted sequential message has to be identical to the tag of the processed message. Is the processed service a concurrent one, the last message may be sequential, but only a call not a return message. All other conditions for state changes are shown in figure 1.

With $mn = ret$ we indicate return messages, with *sequ* sequential and with *conc* concurrent messages. The case of empty output is subsumed under the case of only concurrent output. In the simplest case (sequ-ret) an input call is immediately handled, the stack is not changed. If the output is sequential, the current service is suspended. A concurrent output does not change the stack. The other two cases deal with input return messages, where the stack has to contain an according message invocation, which can be popped (ret-ret) or modified (ret-sequ). In case of modification an according program counter *pc* and an assignment *loc* of local variables denotes the internal state of the service invocation.

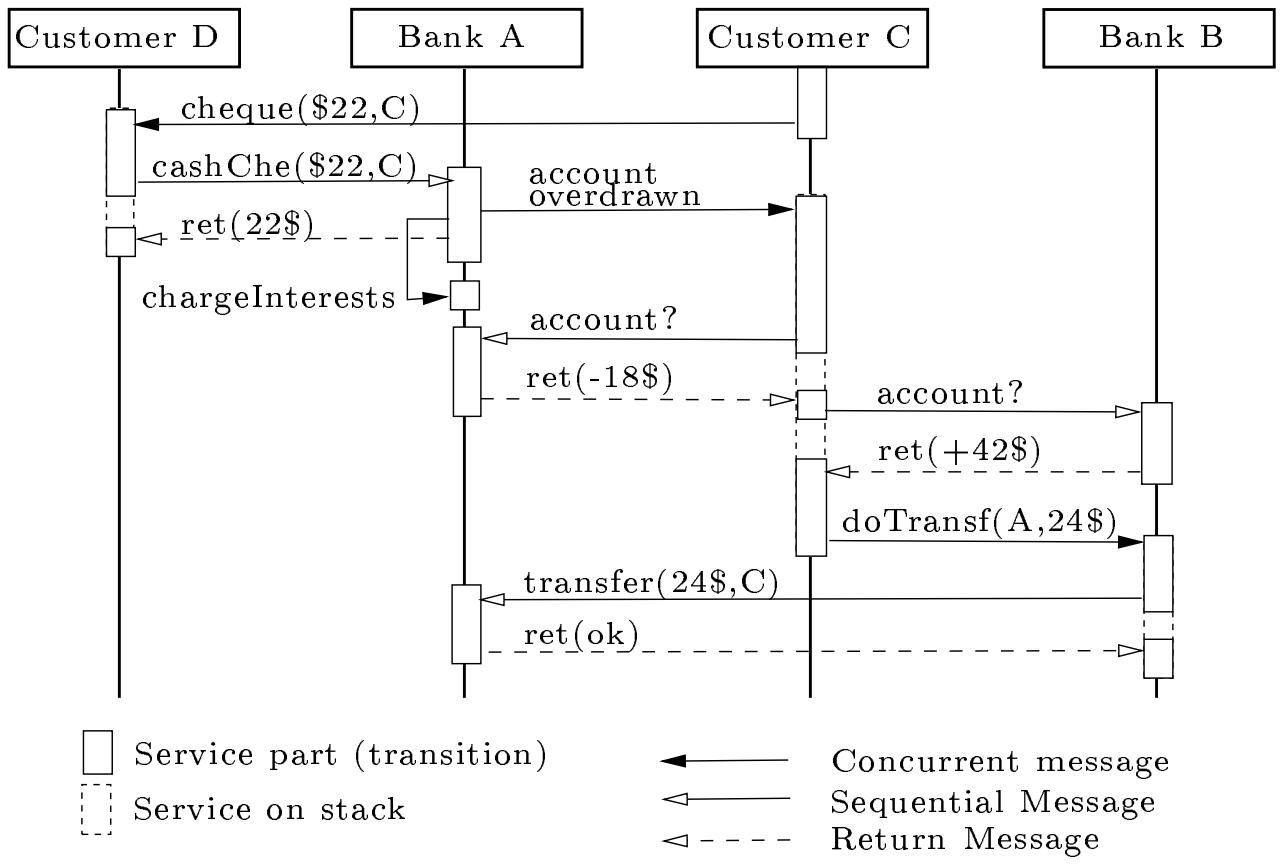

**Figure 2** Bank scenario

We illustrate this model by the following example (see figure 2). Assume we have two customers $C$ and and $D$ as well as two banks $A$ and $B$. Customer $C$ has one account per bank. $B$ gives better interests, but $A$ is used for payment transfers. Customer $C$ uses a cheque for payment of customer $D$. In our concrete scenario, the account in bank $A$ will be overdrawn, after $D$ cashed the check and $C$ gets an according request to balance. Now $C$ is asking for the actual account at both banks and then placing an order to transfer \$24 from bank $B$ to $A$. Bank $B$ awaits the acknowledgment of $A$ before completing the transfer.

## 4 SERVICE DESCRIPTION

In this section, we introduce a state based description technique for services and define object behaviour semantics in terms of $I/O^*$-state machines. We use an abstract version of $I/O^*$-state machines called $I/O^*$-**state transitions diagrams**. They allow for a finite description of the infinite state machines. We use state predicates to partition the state space. Similarly, we allow to abstract from the message parameters by using preconditions referring to attributes and input parameters and by using patterns for input messages. Also, we allow postconditions to describe the effect of data changes and patterns for output messages. The definition given below is a special case of the STD defined in (Grosu, Klein, Rumpe & Broy 1996), where input is restricted to a one-element sequence. Altogether, an $I/O^*$-state transition diagram $(att, I, O, S, \Lambda, \delta, S^0)$ consists of the set $att$ of attributes, the nonempty set $I$ of input messages, the nonempty set $O$ of output messages, a nonempty, finite set of diagram states $S$, a mapping $\Lambda : S \rightarrow \langle Pred \rangle$ associating a predicate over the attributes *att* with the diagram states, a finite transition relation $\delta \subseteq S \times \langle Patt \rangle \times \langle Pred \rangle \times S \times \langle Expr \rangle \times \langle Pred \rangle$, where each transition is labelled with input pattern, precondition, output expression and postcondition, and a nonempty, finite set of initial diagram states $S^0$.

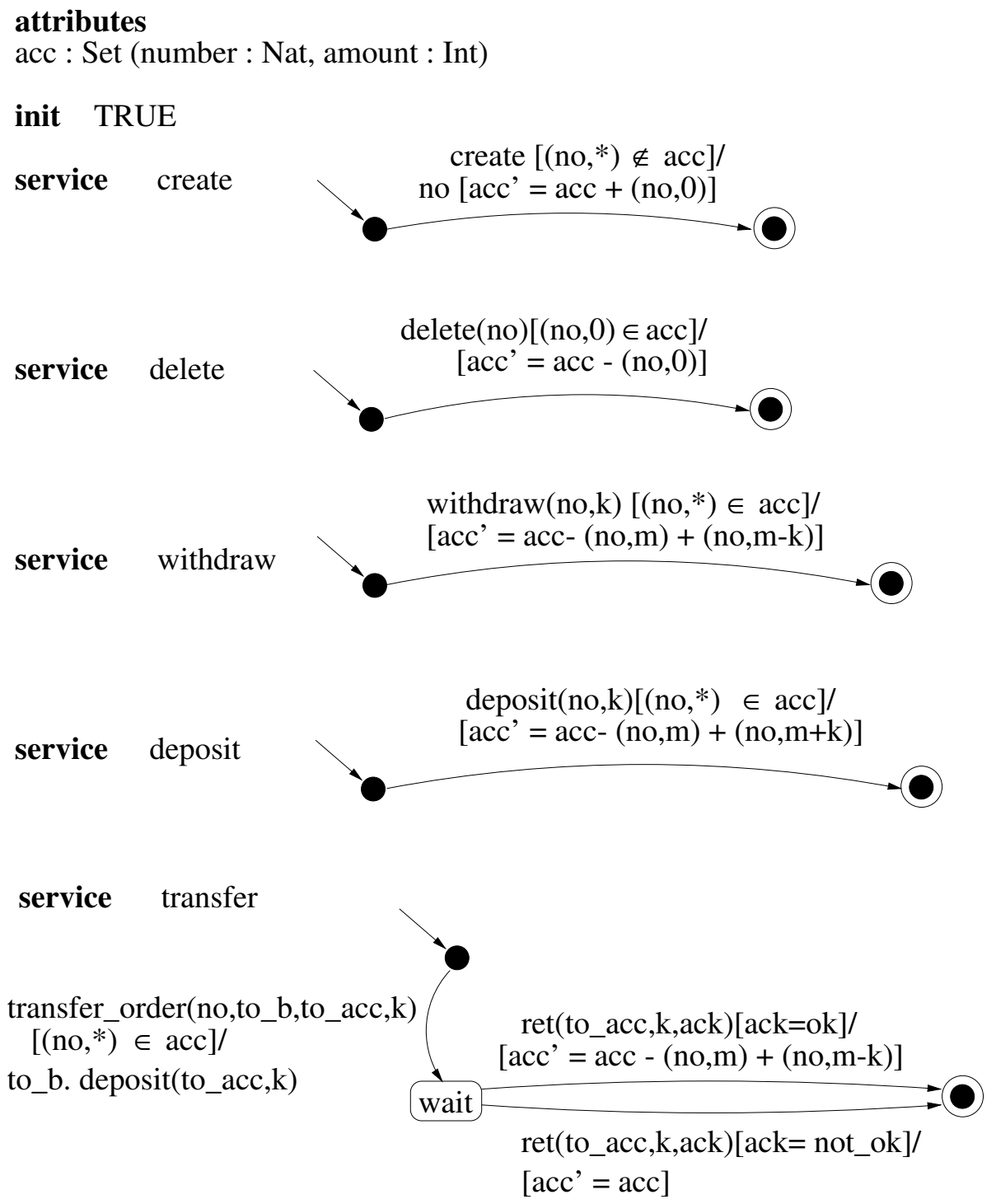

**Figure 3** Bank description with $I/O^*$-STD for each service

$\Lambda(s)$ must be satisfiable for all diagram states $s \in S$ and the predicates of two different diagram states exclude each other. Also the postcondition of a transition must be satisfiable, if the precondition is satisfied.

We call a set of diagrams describing one service each together with a predicate *init* characterizing the initial object states the **object behavior description**. As an example, consider a bank object. Figure 3 shows the object description with attributes defining the state space and with separate service diagrams for `create, delete, withdraw, deposit, transfer`.

The semantics of object behavior description is given in terms of $I/O^*$-state machines. Each diagram transition gives rise to a set of machine transitions satisfying the input pattern, the output pattern and the pre- and postconditions. In addition also the tags and stacks handling the interleaving of services have to be introduced. Thus, let $(att, loc_k, I_k, O, S_k, \Lambda_k, \delta_k, S^0_k)$, $k = 1, ..., n$, be a set of $I/O^*$-STD, where each STD describes a service over the object attributes *att* and the local service variables $loc_k$, and let *init* be a predicate over the attributes. The semantics of this object behavior description is any $I/O^*$-state machine $(\hat{S}, \hat{I}, \hat{O}, \hat{\delta}, \hat{S}^0)$ satisfying the following:

- $\hat{S} = \{\beta \in BEL : \beta.self = id$ and $\beta.tag$ associates with each $tag \in TAG$ a stack of service invocations $SI\}$, where $BEL$ is the set of all variable assignments giving values to the attributes and some additional variables like $self$, tags and for the program counter of the currently active service. The set of service invocations $SI_k$ is given by $(loc_k \rightharpoondown VAL, S_k, ID)$ and $SI = \bigcup_{k=1}^{n} SI_k$. Note that we use the states of the service STD as program counter values.
- $\hat{S}^0 = \{\beta \in \hat{S} : \beta \models init\}^*$
- $\hat{I}$ $(\hat{O})$ is derived from $I$ $(O)$ by using the appropriate message and parameter names and introducing the tag in the messages
- $(\beta_s, (snd_i, rec_i, tt_i, mn_i, ar_i), \beta_t, out ++ (snd_o, rec_o, tt_o, mn_o, ar_o)) \in \hat{\delta}$, if there exists $1 \leq k \leq n$, $T \in \delta_k, \beta \in BEL$ such that $\beta$ satisfies the state predicates, pre- and postconditions, patterns and expressions of $T$ (written as $\beta \models T$) and $\beta|_{att} = \beta_s$ and $\beta|_{att'} = \beta_t'$, where we use the slash notation to denote the values of the variables in the successor state, and either the stack of the tag is empty $(\beta.tag_i = \emptyset)$ and $\beta \models T$ and a new service execution is started $(\beta.pc = s \in S_k^0)$, or the stack is nonempty with program counter $s$ on top $(first(\beta.tag_i) = (\gamma, s, id))$ and $\beta, \gamma \models T$ and the stack is handled according to section 3.

Note that with this semantics the labeling of the diagram states for the services carries a special weight: this labeling describes the set of all states the object may assume while the service is pending at that state. If the state predicate is not satisfied in a state where the pending service is to be continued, arbitrary behavior is possible (due to input enabledness). From a methodological point of view, it sometimes is necessary that services can be guarded from interleaving with other services. For example, account closure should not be possible while transfer is active. This could already be expressed using suitable preconditions and diagram state predicates such that the precondition for account closure is incompatible with the predicate labeling the wait-state of the transfer STD. However, we also allow a more direct way of specification, where diagram states may be labeled with service sets indicating the services which are not allowed to be interleaved at that state (called **exclusion sets**). With this extension, the semantics has to be adopted such that the transitions respect all exclusion sets of pending service invocations $(\pi_4(\pi_2(T)) \notin Ex(u)$ for all $(\gamma, u, id)$ somewhere on some stack* ).

---

*By $\beta \models init$ we denote that formula *init* is satisfied under variable assignment $\beta$

*By $\pi_i$ we select the $i$-th component of a tuple.

## 5 CONCLUSIONS AND RELATED AND FUTURE WORK

We have discussed a semantic model for service execution in the context of multiple threads. We also have introduced a special kind of state transition diagrams for service description and shown how to this object behavior description can be given a precise semantics in terms of state machines taking care of different threads of activity through stacks.

Similar to SDL-92 (Braek & Haugen 1993), services are used to structure object (process) behaviour. In contrast to SDL services, the $I/O^*$-STD description of services makes explicit the state space of the object. This is necessary for an abstract description of service synchronization.

The major difference to Statechart-based description techniques is that we allow services to be distributed over several transitions, while usually only one transition per service is used. The latter kind of modeling is too restrictive, since not all services can be considered to be atomic (e.g. like the transfer service). In Syntropy and O-Mate, for a service additional internal events may be generated. However, a new external event may be treated only when the Statechart has stabilized, that means it has handled all the internal events generated in response to the last external event. Thus, internal events still do not allow e.g. two active transfer services.

Up to now, we have not treated nested states in $I/O^*$-STD. These states are very important for factoring object behavior over orthogonal sets of attributes. Since in our framework we do not allow internal events for communication between different substates, we avoid the usual difficulties of Statechart semantics (von der Beeck 1994). Thus, we do not expect any difficulties with incorporating nested states.

Another point we want to clarify in the near future is the use of refinement techniques as discussed in (Rumpe & Klein 1996). In that paper a calculus of refinement steps on STD is introduced which can be adapted to the framework here without difficulties. We will also explore this notion of refinement as a basis for an inheritance notion covering behavioral properties.

*Acknowledgements*
We thank our colleagues Ursula Hinkel, Peter Scholz and the anonymous referees for helpful comments.

*Bibliography*

**Dr. Barbara Paech** studied Computer Science at the Technical University of Munich, Edinburgh University, and University of Pennsylvania. She received her Ph.D. in Computer Science from the Ludwig-Maximilians-University in Munich. Since 1993 she is a research assistant at the Technical University of Munich where she leads research projects on the formal foundation of software engineering as well as requirements and re-engineering.

**Dr. Bernhard Rumpe** studied Computer Science and Mathematics at the Technical University of Munich. He received his Ph.D. in Computer Science from the Technical University of Munich. His research interests include the formal foundation of state-based behavioral specifications, and functional and OO programming concepts. Since 1997 he leads the SysLab research project on the formal foundation of OO software engineering techniques.